\documentclass[a4paper]{jpconf}
\usepackage{graphicx}
\begin{document}
\title{Emergent gravity: From statistical point of view}

\author{Bibhas Ranjan Majhi}

\address{IUCAA, Post Bag 4, Ganeshkhind,
Pune University Campus, Pune 411 007, India
}

\ead{bibhas@iucaa.ernet.in}

\begin{abstract}
  Near the event horizon of a black hole, the effective theory is two dimensional conformal theory. Here we show that the holographic modes characterising this underlying conformal symmetry and the basic definition of entropy $S$ in statistical mechanics lead the equipartition law of energy. We also show that $S$ is proportional to the gravity action which suggests the emergent nature of gravity. This is further bolstered by expressing the generalised Smarr formula as a thermodynamic relation, $S = E/ 2T$, where $T$ is the Hawking temperature and $E$ is shown to be the Komar energy.
\end{abstract}

\section{Introduction}
   In absence of a true quantum theory of gravity, semi-classical approaches have become increasingly popular and are widely used, particularly in the context of thermodynamics of gravity. It is now evident that gravity and thermodynamics are closely connected to each other \cite{Bekenstein:1973ur,Hawking:1974rv,Bardeen:1973gs}. The repeated failure to quantise gravity led to a parallel development \cite{Jacobson:1995ab,Kothawala:2007em,Verlinde:2010hp}  where gravity is believed to be an emergent phenomenon just like thermodynamics and hydrodynamics instead treating it as a fundamental force. In this context, the fundamental role of gravity is replaced by thermodynamical interpretations leading to similar or equivalent results without knowing the underlying microscopic details. Nevertheless, understanding the entropic or thermodynamic origin of gravity is far from complete since the arguments are more heuristic than concrete and depend upon specific ansatz or assumptions.

   In this writeup, based on our work \cite{Banerjee:2010yd}, we readdressed some of the facts related to the emergent nature of gravity, in the context of the black holes in Einstein gravity.

    Here using certain basic results derived by us \cite{Banerjee:2008sn,Banerjee:2009pf} and applying the standard definition of entropy given in statistical mechanics, we are able to provide a statistical interpretation of gravity. We first give the equipartition law of energy and show that this leads to the identification of entropy with the action for gravity. The immediate consequence of it is that the Einstein equations, obtained by a variational principle involving the action, can be equivalently obtained by an extremisation of the entropy. This implies that the gravity can be thought of as the emergent phenomenon.

     The emergent nature has been further bolstered by deriving the relation $S=E/2T$, connecting the entropy ($S$) with the Hawking temperature ($T$) and energy ($E$) for a black hole with stationary metric.  We show that this energy corresponds to Komar's expression \cite{Komar:1958wp,Wald:1984rg}. Using this fact we show that the relation $S=E/2T$ is also compatible with the generalised Smarr formula \cite{Smarr:1972kt,Bardeen:1973gs,Gibbons:1976ue}.

\section{Previous results}
  Here we briefly introduce three relevant results, which were derived previously by some of us, for our main purpose.
\vskip 2mm
\noindent
$\bullet$ Thermodynamics of a black hole is universally governed by its properties near the event horizon. It is also well understood that near the event horizon the effective theory becomes two dimensional conformal theory whose metric is given by the two dimensional ($t-r$)- sector of the original metric. For details see \cite{Majhi:2011yi}.
\vskip 2mm
\noindent
$\bullet$  Using the WKB approximation, the left ($L$) and right ($R$) moving (holomorphic) modes are obtained by solving the appropriate field equation under the effective two dimensional metric. It has been shown that the modes inside and outside the horizon are related by the transformations \cite{Banerjee:2008sn,Banerjee:2009pf}: 
\begin{eqnarray}
\phi^{(R)}_{in} = e^{-\frac{\pi\omega}{\kappa}} \phi^{(R)}_{out}; \,\,\
\phi^{(L)}_{in} = \phi^{(L)}_{out}
\label{mode1}
\end{eqnarray}
where ``$\omega$'' is the energy of the particle as measured by an asymptotic observer and ``$\kappa$'' is the surface gravity of the black hole.
Furthermore, the $L$ mode gets trapped while the $R$ mode tunnels through the horizon and is observed at asymptotic infinity as Hawking radiation \cite{Banerjee:2008sn,Banerjee:2009pf}.
The probability of this ``$R$'' mode, to go outside, as measured by the outside observer is given by 
\begin{eqnarray}
P^{(R)}=\Big|\phi^{(R)}_{in}\Big|^2 = \Big|e^{-\frac{\pi\omega}{\kappa}} \phi^{(R)}_{out}\Big|^2 = e^{-\frac{2\pi\omega}{\kappa}}
\label{mode3}
\end{eqnarray} 
where, in the second equality, (\ref{mode1}) has been used. This is essential since the measurement is done from outside and hence $\phi^{(R)}_{in}$ has to be expressed in favour of $\phi^{(R)}_{out}$.
\vskip 2mm
\noindent
$\bullet$ The effective two dimensional curved metric can be embedded in a flat space which has exactly two space-like coordinates. This is a consequence of a modification in the original GEMS (globally embedding in Minkowskian space) approach of \cite{Deser:1998xb} and has been elaborated by us in \cite{Banerjee:2010ma}. This tells that each $R$ mode can be associated to two degrees of freedom. Therefore, the total number of degrees of freedom for $n$ number of $R$ modes is $N=2n$.

\section{Partition function}
  The partition function for the space-time with matter field is given by  \cite{Gibbons:1976ue},
\begin{eqnarray}
{\cal{Z}} = \int ~ D[g,\Phi] ~  e^{i I[g,\Phi]}
\label{1.01}
\end{eqnarray}
where $I[g,\Phi]$ is the action representing the whole system and $D[g,\Phi]$ is the measure of all field configurations ($g,\Phi$).  
Since we want to confine ourself within the usual semi-classical regime, we shall neglect all the higher order terms for the subsequent analysis.
Therefore, in the semi-classical regime the partition function is expressed as \cite{Gibbons:1976ue},
\begin{eqnarray}
{\cal{Z}} \simeq e^{i I[g_0,\Phi_0]},
\label{1.04}
\end{eqnarray}
where ($g_0,\Phi_0$) are the background fields and the classical action $I[g_0,\Phi_0]$ leads to the Einstein equation.

\section{Relation between different thermodynamical quantities with the action and their implications}
  Here using the previous results and the above definition of partition function combined with the standard definition of entropy in statistical mechanics we derive three important relations among different thermodynamical quantities of a stationary black hole. The physical implications are also being described at the end.

 In statistical mechanics the entropy is related to the partition function by the relation:
$S = \ln{\cal{Z}} + \frac{E}{T}$ and so (\ref{1.04}) leads to,
$S= i I[g_0,\Phi_0] + \frac{E}{T}$
where $E$ and $T$ are respectively the energy and temperature of the system. Now since there are $N$ number of degrees of freedom in which all the information is encoded, the entropy ($S$) of the system must be proportional to $N$. Hence
\begin{eqnarray}
N = N_0 S = N_0 (i I[g_0,\Phi_0]  + \frac{E}{T}),
\label{1.06}
\end{eqnarray} 
where $N_0$ is a proportionality constant. Now the average value of the energy, measured from outside, is calculated as,
\begin{eqnarray}
<\omega> = \frac{\int_0^{\infty}~d\omega~\omega ~P^{(R)}}{\int_0^{\infty}~d\omega~ P^{(R)}} = T
\label{mode4}
\end{eqnarray}
where $T=\kappa/2\pi$ is the temperature of the black hole \cite{Banerjee:2009pf}. Therefore if we consider that the energy $E$ of the system is encoded near the horizon and the total number of pairs created is $n$ among which this energy is distributed, then we must have,
$E=nT$
where only the $R$ mode of the pair is significant.
Hence, since $N=2n$, we obtain the energy of the system as
\begin{eqnarray}
E = \frac{1}{2}NT.
\label{1.07}
\end{eqnarray}
Noted that (\ref{1.07}) can be interpreted as the usual {\it law of equipartition of energy}, since it implies that if the energy $E$ is distributed equally over each degree of freedom, then 
 each degree of freedom should contain an energy equal to $T/2$. The fact that the energy is equally distributed among the degrees of freedom may be understood from the symmetry of two space-like coordinates ($z^1\longleftrightarrow z^2$) such that the metric is unchanged \cite{Banerjee:2010ma}. 

 Now using (\ref{1.06}) and (\ref{1.07}) we obtain following relations:
\begin{eqnarray}
E=\frac{N_0}{2 - N_0}iT I[g_0,\Phi_0]; \,\,\,\
S=\frac{2E}{N_0T}.
\label{entropy1}
\end{eqnarray}
In order to fix the value of ``$N_0$'' we consider the simplest example, the Schwarzschild black hole for which the entropy, energy and temperature are given by,
$S = \frac{A}{4} = 4\pi M^2, \,\,\  E = M,  \,\,\ T = \frac{1}{8\pi M}$,
where ``$M$'' is the mass of the black hole.
Substitution of these in the second relation of (\ref{entropy1}) leads to $N_0 = 4$.
Finally, putting back  $N_0=4$ in the preceding relations we obtain,
\begin{eqnarray}
E = -\frac{i\kappa I[g_0,\Phi_0]}{\pi}; \,\,\
S=-iI[g_0,\Phi_0] = \frac{E}{2T}.
\label{entropy}
\end{eqnarray}
The implications of the above relations are as follows:
\vskip 2mm
\noindent
$\bullet$ Use of the explicit form of Einstein-Hilbert action in the first relation shows that $E$ is the Komar
conserved quantity corresponding to Killing vector. For details, see \cite{Banerjee:2010yd}.
\vskip 2mm
\noindent
$\bullet$ First equality of the second relation signifies that the extremisation of entropy leads to Einstein's equations. It illustrates the emergent nature of gravity.
\vskip 2mm
\noindent
$\bullet$ The last equality $S=E/2T$ is the general expression for Smarr formula. This can be checked by substituting the relevant quantities of Kerr-Newman black hole to obtain the relation 
$\frac{M}{2} = \frac{\kappa A}{8\pi} + \frac{VQ}{2} + \Omega J$, which is 
the generalised Smarr formula \cite{Smarr:1972kt,Bardeen:1973gs,Gibbons:1976ue}.
This further clarifies the fact that the gravity is an emergent phenomenon.

\section{Conclusions}

    We, in the context of black hole, using the statistical definition of entropy and the partition function for gravity, further clarified the possibility of considering gravity as an emergent phenomenon. There are certain issues still to be addressed or clarified. Let us tabulate some of them: 
\vskip 2mm
\noindent
$\bullet$ First of all, we saw that the proportionality constant $N_0$ was not fixed from the fundamental ground. Rather, we did it by using the parameters of a specific black hole. An independent derivation will be very much interesting.
\vskip 2mm
\noindent
$\bullet$ Is it possible to discuss the same for more general situation, like the spacetime without black
hole? This may be done by considering the local Rindler frame in the spacetime  which has a null surface on which the temperature and entropy can be associated.
\vskip 2mm
\noindent
$\bullet$ A more simpler but interesting extension would be inclusion of higher dimensions. In this regard, one can be interested to see if the higher dimensional gravity exhibit same feature or in other words if the above features depend on dimension of spacetime. Some part of this issue has already been addressed in \cite{Banerjee:2010ye}. Here we did for the higher dimensional black hole in Einstein gravity and showed that we again have a relation similar to $E=2ST$, but $E$ is interpreted as the Komar conserved quantity rather than the energy itself.
\vskip 2mm
\noindent
$\bullet$ As a final remark we feel that although our results were derived for Einstein gravity, the methods are general enough to include other possibilities like $f(R)$ gravity, Lancozs-Lovelock gravity etc.

\end{document}